\documentclass[twocolumn,secnumarabic, graphics,floatfix,nofootinbib,nobibnotes,aps,prb,10pt]{revtex4-1}
\usepackage[english]{babel}
\usepackage{epsf}
\usepackage{epsfig}
\usepackage{graphicx}
\usepackage{xcolor}
\usepackage[normalem]{ulem} 

\begin{document}

\title{Coherent Tunneling and Strain Sensitivity of an All$-$Heusler Alloy Magnetic Tunneling Junction: A First-Principles Study}

\author{Joydipto Bhattacharya$^{1,2}$, Ashima Rawat$^{3}$, Ranjit Pati$^{3}$, Aparna Chakrabarti$^{\dag}$$^{1,2}$, Ravindra Pandey$^{3}$}

\affiliation{$^{1}$Raja Ramanna Centre for Advanced Technology, Indore 452013, India}
\affiliation{$^{2}$Homi Bhabha National Institute, Training School Complex, Anushakti Nagar, Mumbai 400094, India}
\affiliation{$^{3}$Department of Physics, Michigan Technological University, Houghton, Michigan 49931, USA}

\begin{abstract}
Half-metallic Co-based full Heusler alloys have captured considerable attention of the researchers in the realm of spintronic applications, owing to their remarkable characteristics such as exceptionally high spin polarization at Fermi level, ultra-low Gilbert damping, and high Curie temperature. In this comprehensive study, employing density functional theory, we delve into the stability and electron transport properties of a magnetic tunneling junction (MTJ) comprising a Co$_2$MnSb/HfIrSb interface. 
Utilizing a standard model given by Julliere, we estimate the tunnel magnetoresistance (TMR) ratio of this heterojunction under external electric field, revealing a significantly high TMR ratio ($\approx500\%$) that remains almost unaltered for electric field magnitudes up to 0.5 V/\AA. In-depth investigation of K-dependent majority spin transmissions uncovers the occurrence of coherent tunneling for the Mn-Mn/Ir interface, particularly when a spacer layer beyond a certain thickness is employed. Additionally, we explore the impact of bi-axial strain on the MTJ by varying the in-plane lattice constants between -4$\%$ and +4$\%$. Our spin-dependent transmission calculations demonstrate that the Mn-Mn/Ir interface manifests strain-sensitive transmission properties under both compressive and tensile strain, and yields a remarkable three-fold increase in majority spin transmission under tensile strain conditions. These compelling outcomes place the Co$_2$MnSb/HfIrSb junction among the highly promising candidates for nanoscale spintronic devices, emphasizing the potential significance of the system in the advancement of the field.

\end{abstract}
\maketitle

\section{Introduction}
In recent years, significant advancements have been made in controlling spin-dependent tunneling between two ferromagnetic electrodes separated by an insulating barrier. These developments have a profound impact on various magnetic data storage technologies, particularly due to the observation of exceptionally high tunneling magnetoresistance (TMR) values\cite{Ikeda,Ralph,Grollier2020}. Initially, the ability to achieve substantially high TMR in magnetic tunneling junctions (MTJs) was limited by the use of amorphous tunnel barriers. However, the landscape has since evolved significantly, thanks to the theoretical predictions\cite{Valev_05,Miura,Miura11,Peter} and subsequent experimental realization of epitaxial MTJs\cite{Sakuraba_2005,Yuasa_2007}.\\

The fabrication of epitaxial Co-based MTJ, which exploits the coherent electronic tunneling phenomenon to produce large TMR, was a major breakthrough in the field of MTJs. TMR values of up to 220$\%$ at room temperature and 300$\%$ at low temperatures for CoFeB/MgO/CoFeB based MTJ was reported by Parkin and his co-workers in 2004.\cite{Parkin2004} Till now, the highest TMR ratio in Heusler alloy based MTJs has been observed for Co$_2$MnSi/MgO/Co$_2$MnSi junction, which produced a TMR ratio of 1995$\%$ at 4K \cite{Liu12} which reaches up to 2610$\%$ with Mn-rich and highly Fe-doped electrode.\cite{Liu_2015} However, from ab-initio theory based calculations, TMR ratio of about 10$^5 - 10^8$  has been reported for MTJs with half-metallic electrodes.\cite{Feng_2022,Rotjanapittayakul,Zhou_2021,Balcı2019,Lu_2021,Larionov_2022,Paudel_2019} \\

In this context, the half-metallic (HM) materials, exhibiting metallic behavior for one of the spin-up and spin-down channels and semiconducting for the other, have long been expected to work well as a spin filter or spin-injecting source capable of generating extremely highly spin-polarized current.\cite{Ma_2017,Fusil_2006} Among the various HM materials that have been explored, HM Heusler alloys are regarded as one of the most promising materials for the electrode, due to their low Gilbert damping factor, high Curie temperature and reasonably good lattice matching with the traditionally used semiconductor substrates, $e.g.$ MgO, GaAs etc.\cite{Miura,Miura11,Peter} Since the prediction of first HM Heusler alloy by de Groot $et.al$ \cite{Groot_1983}, many HM Heusler alloy materials have been proposed from first-principles calculations\cite{LUO2008421,Picozzi_2002,Khandy_2020,Faleev_2017} and many of these materials were discovered experimentally as well.\cite{RANI2019165662,Mahat_2021,Luo_2007} This gives us a wide range of materials to choose from, suitability depends on their electronic and geometric properties. \\

These HM materials being close to 100$\%$ spin-polarized at Fermi level (E$_F$), provide an enormous advantage over other ferromagnetic electrode materials, leading to its wide application in spintronic devices. However, interestingly, most of the half metals loose their unique character ($i.e$ $\sim$100$\%$ spin polarization at E$_F$), when they are embedded into heterostructures constructed for the purpose of achieving a high TMR ratio or efficient spin injection into semiconductor spacer layers.\cite{Nagao_2006} So far, there are quite a few studies on the electronic properties of heterojunction interfaces based on first-principles calculations.\cite{Miura11,Miura,Peter,BHATTACHARYA_2023} It is known that HM properties almost always get affected and completely lose the spin-polarized character at interfaces. However, there are a few theoretical exceptions. For example, NiMnSb/CdS, zinc-blende CrAs/GaAs and all Heusler interfaces (Co$_2$MnSi/Fe$_2$TiSi, Co$_2$MnSb/TiCoSb,  CoFeTiSi/Fe$_2$TiSi.\cite{BHATTACHARYA_2023,Feng_2022,Wijs_2001,Nagao_2004}. It is still not well understood how much the interfacial electronic structures affect the spin polarization. As long as the functioning of devices relies on coherent tunneling, their performance seems to be sensitive to the condition of the interfaces.\cite{Butler_2008,BHATTACHARYA_2023}\\

The Heusler alloy family is particularly attractive for constructing spintronic devices due to the wide range of electronic properties it offers, along with its suitable geometry. Recent literature has highlighted the potential for achieving high TMR values and interface spin polarization at E$_F$ (henceforth denoted by SP) in all-Heusler based magnetic tunneling junctions at low bias. Several studies have reported promising findings.\cite{Gopikrishnan,Feng_2022,BHATTACHARYA_2023} In our present investigation, as in recent past\cite{BHATTACHARYA_2023}, we have probed Co$_2$MnSb, a half-metallic full-Heusler alloy, as an electrode in spintronic devices. This alloy demonstrates exceptional bulk properties, including a high Curie temperature, low Gilbert damping, and defect tolerance, all of which are essential for successful applications in spintronics. \\

 Although MTJ based magnetoresistive random access memory (MRAM) is quite area efficient, there have been intensive challenges that needed to be overcome as far as the technology is concerned. One of the key challenges is to reduce the writing current density, while maintaining the low resistance area (RA) product. So, to maintain a reasonably high TMR ratio and low RA product (by reducing the barrier height at the interface), it is therefore crucial to explore various material combinations. This can offer more flexibility and freedom to engineer the device properties, by tuning the interfaces and expanding the search for material combinations beyond those that have already been adopted by the community.\\
 
Here, for the spacer material, we are focusing on the family of Heusler alloys with 18 valence electron, per unit cell, specifically ternary XYZ compounds. These compounds have elements at X, Y, and Z sites in a 1:1:1 stoichiometry and consist of diverse chemical groups with atoms spanning columns 1-5, 9-17 in the Periodic Table, providing a vast range of materials. However, not all possible combinations result in stable compounds. Out of 483 possible combinations, only 83 have been experimentally synthesized. By using ab-initio based electronic structure calculations on the remaining 400 unreported compounds, it is predicted that 50 more compounds should be stable.\cite{Yan2015,Gautier2015} Further, based on these predictions, 15 compounds were experimentally grown. The XYZ compounds with heavy elements like Sb, Bi, Sn, or Pb occupying the Z site tend to have a stable cubic structure, whereas those with light Z atoms (O, S, Se, N, P, As, C, Si, and Ge) tend to have non-cubic structures and are often metallic \cite{Gautier2015,Yan2015,MBPRB2019}. Therefore, out of the 108 experimentally synthesized 18 valence electron compounds, we must select a material that is formed in a cubic phase with a relatively low lattice mismatch with the chosen electrode. The spacer material should also exhibit a relatively low direct bandgap (around 1-2 eV) at $\Gamma$, which is crucial in achieving a low resistance-area (RA) product and highly coherent tunneling. As the spacer material, we have thus chosen the direct bandgap semiconductor, HfIrSb. This semiconducting half-Heusler alloy exhibits the lowest decay rate at the $\Gamma$ point and has recently been experimentally synthesized in a single phase.\cite{Yan2015}\\

Strain engineering is another alternative method to enhance the writing power, which has been demonstrated for CoFe-based leads, showing possibilities to modify TMR ratio and magnetocrystalline anisotropy.\cite{Loong2014,Roschewsky} Several studies have been proposed for the strain mediated switching mechanisms to lower the energy barrier between parallel (P) and antiparallel (AP) states.\cite{Biswas2017} It has been shown in earlier studies that strain can control the half-metallicity, spin polarization at E$_F$ and magnetic moments.\cite{Kanomata_2010,wu2019elastic} 

In this work, we propose a combination of full and half Heusler-based MTJ with an interface derived by mixing of alloys. This new all-Heusler based junction (Co$_2$MnSb/HfIrSb) possesses smoothly varying electronic properties, ranging from half-metallic and  magnetic electrode, to semiconducting and nonmagnetic spacer.  To determine which stacking sequence conserves the high SP, we perform electronic structure calculations with various interface geometries. Our calculations suggest that the HM property is maintained for the Mn-Sb as well as (Mn-rich) Mn-Mn (001) surfaces of the electrode material. Further, we investigate the spin transport properties for heterojunctions with Mn-Sb/Ir and Mn-Mn/Ir interfaces and find the tunneling to be highly coherent. Additionally, we explore the influence of bi-axial strain on the spin-dependent transmission which reveals that the introduction of controlled mechanical strain has a profound impact on the quantum transport behavior of MTJs. Specifically, we show that the majority spin transmission can be enhanced up to threefold due to the modified electronic structure of the junction induced by the applied strain. These results have important implications for the design and optimization of high-performance MTJs \cite{Noh_19,Loong2014,Zhao_2016}.

The paper is organized as follows. After a brief description of the methodology adopted in the paper, we first discuss about the electronic and magnetic properties of the bulk electrode (Co$_2$MnSb) and spacer (HfIrSb) materials by explaining the criteria that brought us to focus in this specific combination. Thereafter we discuss the electronic and transport properties of the heterostructures with different interfaces. Then we show the strain-sensitive transmission properties of the most energetically stable interface (Mn-Mn/Ir) under both compressive and tensile strain. Finally we conclude our work.

\section{Methodology}
To perform electronic structure calculations, we utilize the density functional theory (DFT) based Vienna Ab-initio Simulation Package (VASP)\cite{Kresse}, using the projector augmented wave method implemented in VASP.\cite{Kresse99} We employ the Perdew-Burke-Ernzerhof\cite{Perdew} generalized gradient approximation (GGA) for the exchange-correlation (XC) functional. An energy cut-off of 500 eV has been used to expand the plane waves. The Brillouin zone was sampled using the Monkhorst-Pack scheme\cite{Monkhorst} with a mesh of 17$\times$17$\times$1 k-points. A convergence criterion for energy in the self-consistent-field cycle of 10$^{-6}$ eV is adopted. To optimize the geometry of the heterojunction systems, we fix the in-plane lattice constant of the electrode material. The total force tolerance on each atom is set to be below 0.02 eV/\AA. For calculating the density of states, we use the tetrahedron method of integration scheme, implemented in the VASP package, with a mesh of 21$\times$21$\times$1 k-points.
	
	The value of spin polarization at E$_F$ (SP) has been calculated as follows:\\	
	\begin{center}
		$SP = \frac{n^{\uparrow}(E_F)- n^{\downarrow}(E_F)}{n^{\uparrow}(E_F)+ n^{\downarrow}(E_F)}\times 100 \% $
	\end{center}
	Here n$^\uparrow(E_F)$ and n$^\downarrow(E_F)$ correspond to the majority and minority spin density of states (DOS) at E$_F$, respectively. \\
 
To investigate the magnetoelectric (ME) property and to understand the effect of spin injection bias on the TMR ratio, we have incorporated transverse electric field in our calculations. We have considered series of  external electric
fields ranging from	0.01 V/\AA~ to 0.50 V/\AA ~in the direction perpendicular to the heterojunction interfaces. For the electric field dependent calculation 15 \AA~ of vacuum has been added along the (001) direction with the same in-plane lattice parameter as obtained in the previous interface. To mitigate any artificial Coulomb interaction resulting from the external electric field, a dipole correction has been incorporated.\\

In order to calculate the TMR ratio for each applied electric field, we have taken the help of the model given by Julliere, which is based on two-current model\cite{JULLIERE}. 

Since, this model assumes that spin is conserved in tunneling in a MTJ, the TMR ratio can be calculated as follows:
\begin{equation}
TMR=\frac{G_P -G_{AP}}{G_{AP}}
\end{equation}
where G$_P$ and G$_{AP}$ can be calculated from the projected density of states (PDOS) of the bottom and top magnetic contact in both parallel (P) and anti-parallel (AP) configurations as given by,

\begin{equation}
G_P = \frac{e^2}{h}(n^{\uparrow}_{bottom}n^{\uparrow}_{top} +n^ {\downarrow}_{bottom}n^{\downarrow}_{top} )
\end{equation} 

\begin{equation}
G_{AP} =\frac{e^2}{h}(n^{\uparrow}_{bottom}n^{\downarrow}_{top} +n^ {\downarrow}_{bottom}n^ {\uparrow}_{top} )
\end{equation} 

where, $ n^{{\uparrow}/{\downarrow}}_{bottom}$ and $ n^{{\uparrow}/{\downarrow}}_{top}$ represent the PDOS value of majority/minority spin carrier for bottom and top magnetic contact at E$_F$, respectively and $e$ and $h$ are the electronic charge and Plank's constant respectively.
 
	Furthermore, the first-principles calculations of ballistic conductance have been carried out using PWCOND code \cite{Smogunov_04} as implemented in the Quantum ESPRESSO (QE) package\cite{Baroni}. After obtaining the converged geometry using VASP, we calculate the spin-dependent transmission,  $T^{\sigma}(k_{||},E)$ using the method proposed by Choi and Ihm\cite{Choi}, also using GGA exchange-correlation functionals\cite{Perdew}. The spin-dependent tunneling conductance is obtained by using Landauer-Büttiker  formula \cite{Smogunov_04}:\\
	\begin{center}
		$G^{\sigma} = \frac{e^2}{h}\sum_{k_{||}}T^{\sigma}(k_{||},E_F)$
	\end{center}
	where, $\sigma (= \uparrow,\downarrow$), is the spin index and $T^{\sigma}(k_{||},E_F)$ is the spin-dependent transmission coefficient at the energy $E_F$, with $k_{||} = (k_x, k_y)$. We set the wave function and charge density cut-off energy to 60 and 600 Ry, respectively, and use a 10$\times$10$\times$1 k-point mesh for the heterojunction calculations. All calculations are converged to an accuracy of 10$^{-8}$ Ry. We resolve the transmission with a large k-grid in the $x$ and $y$ directions (100 $\times$ 100) to accurately capture fine spikes in transmission. To reduce the 2D plane wave basis set, we use an energy window of 45 Ry. More information about the method for calculating ballistic conductance can be found in Ref.\onlinecite{Choi}. \\

The Crystal Orbital Hamilton Population (COHP) analysis has been carried out using the LOBSTER package \cite{Deringer2011,Dronskowski1993}.The pbeVaspFit2015 basis with the following basis functions:
Co: 3d, 4s, and 3p; Mn: 4s, 3d and 4p; Sb: 5s, 4d and 5p; Hf: 5d, 6s and 5p ; Ir: 5d, 6s and 4f, have been used for the orbital projection of plane waves. The wavefunctions are obtained from the DFT calculations.
	
\section{Results $\&$ Discussion}

\subsection{Electronic and Magnetic properties of Bulk and Surfaces (001) of Co$_2$MnSb Alloy}
The HM nature of the bulk is confirmed through comprehensive electronic structure calculations, as presented in Table S1 and Fig. S1.\cite{supply} Analysis reveals the presence of $\Delta_1$ and $\Delta_5$ symmetric bands within the majority spin channel along the transport direction (from $\Gamma$ to X shown in Fig. S1\cite{supply}). These bands play a crucial role in facilitating efficient spin-dependent symmetry filtering transport. Supplementary Information\cite{supply} gives details of the electronic properties of the bulk electrode. \\

We further investigate the impact of bi-axial strain on the electronic and magnetic properties of the bulk electrode. Tensile strain preserves the half-metallic (HM) behavior, while compressive strain destroys it (Table S2, Fig. S2 \cite{supply}). The band dispersion and orbital character along the transport direction remain largely unaffected by strain (Fig. S3, S4, S5 \cite{supply}). However, under compressive strain, we observe the appearance of $\Delta_1$ symmetric bands in the minority spin states (Fig. S3 \cite{supply}), likely to influence the transmission behavior, which will lead to the reduction of  TMR ratio under compressive strain. 
 
To investigate the interfacial properties of the ferromagnetic/semiconductor heterostructure, we initially examined the free-standing surface slabs of Co$_2$MnSb (001). Three different atomic terminations were considered: Co-Co, Mn-Sb, and Mn-excess Mn-Mn, each consisting of 17 diatomic layers with a 15 \AA~ vacuum along the z direction to prevent any interaction between the slabs along the (001) direction due to the periodic arrangement of the same. The energetic, electronic, and magnetic properties of these surfaces are presented in Table S3.\cite{supply} Analysis of the projected density of states confirmed the preservation of the HM property in case of the Mn-Sb and Mn-Mn terminated surfaces (Fig S8 \cite{supply}). As Heusler alloys with high surface SP are desirable for spintronic devices, we selected these two surface terminations for further investigations on the heterostructures with semiconductors. \\

\subsection{Bulk properties of HfIrSb Alloy}
In the MgAgAs-type crystal structure of HfIrSb, there are four interpenetrating FCC sublattices: a rock-salt structure formed by a lattice of Hf atoms and a lattice of Sb atoms, and a lattice of Ir atoms occupying the center of every other Hf4Sb4 cube. The remaining Hf4Sb4 cubes have vacant centers. The calculated lattice constant and band gap of HfIrSb (Table S1\cite{supply}) agree well with other calculations performed with the GGA XC term.\cite{Lee_2011,Arikan2020} The lattice constant of HfIrSb is about 5$\%$ larger than that of Co$_2$MnSb. This difference has implications for the local bonding between atoms, as well as for the resulting electronic and transport properties, which will be discussed later. \\

\begin{figure}
	\begin{center}
		\includegraphics[width=.5\textwidth]{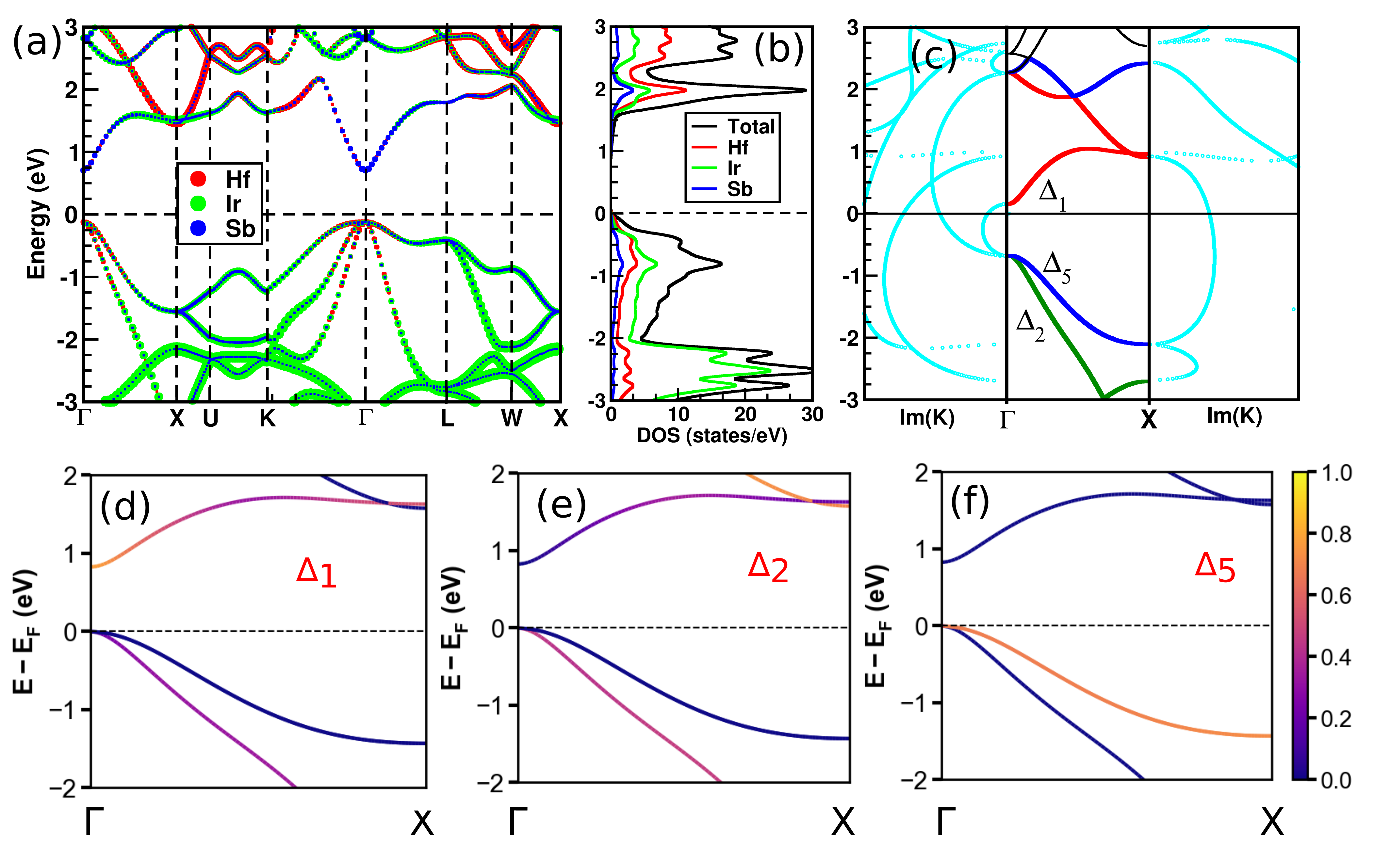}
		\caption{The electronic properties of the HfIrSb spacer material are presented. Panel (a) shows the atom-projected band structure, while panel (b) shows DOS. Panel (c) depicts the complex band structure along the $\Gamma$ to X direction. Finally, panels (d)-(f) display the orbital-projected band structures. Here $\Delta_1$ , $\Delta_2$ , $\Delta_5$ represent the d$_{z^2}$, (d$_{x^2 - y^2}, d_{xy}$) and (d$_{xz}, d_ {yz})$ orbital characters, respectively.}		
		\label{Fig:1}		
	\end{center}
\end{figure}

Fig. \ref{Fig:1} depicts the atom projected band structure along the high symmetry path in the 1st Brillouin zone, along with the total DOS for HfIrSb. The compound has a direct bandgap of approximately 0.89 eV at the $\Gamma$ point, whereas the experimental band gap is found to be 1.3 eV.\cite{Lee_2011} The atom-projected band structure suggests that the top of the valence band is mainly composed of contributions from Hf and Ir atoms, whereas the conduction band is predominantly due to Sb atoms. Since Hf and Ir have $d$ electrons in the valence, the valence band is expected to be mostly $d$ electron derived.\\

 Further, it is crucial to understand the symmetries of the conduction band minima (CBM) and valance band maxima (VBM) of the spacer material because we expect that the contribution of electrons for transport with different symmetries also differs.  As can be seen from the orbital projected band-structure in Fig. \ref{Fig:1}(d) to (f), there are two valence bands  along the  $\Gamma$ to X direction that touches the E$_F$. One has predominantly  $\Delta_1$ and $\Delta_2$ orbital symmetry (Fig. \ref{Fig:1}(d), (e)) whereas the other shows predominant $\Delta_5$ orbital character (Fig. \ref{Fig:1} (f)). However, due to the presence of $sp$ electrons in the valence shell of Sb atoms, the CBM exhibits dominant $\Delta_1$ symmetry. This suggests both the  $\Delta_1$ and $\Delta_2$ states are likely to contribute to tunneling. We extend our investigation to examine the impact of spin-orbit coupling (SOC) interaction on the electronic band structure of HfIrSb. In Figure S6 \cite{supply}, the band structure of HfIrSb with SOC is presented. The introduction of SOC breaks the degeneracy of the valence bands at the $\Gamma$ point, leading to a reduction of the band gap to 0.67 eV. Furthermore, we observe the splitting of valence bands as we move away from $\Gamma$. Notably, conduction bands near the $\Gamma$ point exhibits a Zeeman-like spin splitting of bands. Analyzing the orbital-projected bands (Figure S6 (b) - (d) \cite{supply}) incorporating SOC, we discern that while the CBM is predominantly of $\Delta_1$ orbital character, the valence band maximum (VBM) is primarily characterized by the $\Delta_5$ orbital.  \\

The periodicity of the bulk crystal requires that the Bloch k-vectors are real, but in metal-semiconductor (or metal-insulator) interfaces, the metal-induced gap states (MIGS) play a crucial role. These states are itinerant in the metallic electrodes and exponentially decaying in the insulator. Their solutions with complex-k vectors result in complex band structures for the insulators.\cite{DEDERICHS2002108} In this study, we compare the complex band structures of HfIrSb along the (001) direction, which is the propagation direction of electrons in the heterojunctions.

To investigate the tunneling behavior in HfIrSb, we perform band structure calculations of the spacer material. Both real and complex K-values at k$_{||}$ = $\Gamma$ were considered, as depicted in Fig. \ref{Fig:1}(c). Understanding the decay rate in the barrier layer, where the complex band energies intersect with the E$_F$, represented by Im(k), is crucial for comprehending how tunneling electrons approach the barrier layer perpendicular to the surface. A lower decay rate suggests that the electrons travel a shorter effective distance before encountering the barrier layer \cite{DEDERICHS2002108}. For HfIrSb, we observe that the complex bands intersected at the E$_F$ at a specific k-point in the complex region, resulting in a lower decay rate of Im(k) = 0.11$\frac{2\pi}{a}$ at the $\Gamma$ point (Fig.~\ref{Fig:1}(c)), in contrast to the case of MgO (0.21$\frac{2\pi}{a}$).\cite{BHATTACHARYA_2023} In our previous study on a MTJ with TiCoSb as spacer layer, which is an indirect band gap semiconductor, we have found decay rates of 0.25$\frac{2\pi}{a}$ at the $\Gamma$ point and 0.14$\frac{2\pi}{a}$ at the X point.\cite{BHATTACHARYA_2023} Therefore, we anticipate a larger $\Gamma$-centric tunneling  in case of HfIrSb spacer layer compared to TiCoSb and MgO\cite{Butler_2008}.

\subsection{Co$_2$MnSb/HfIrSb/Co$_2$MnSb Heterojunction}
\subsubsection{Electronic and Magnetic properties of the Heterojunction}	

\textbf{Geometric Structure and Stability:}
In this section, we examine the electronic and magnetic properties of Co$_2$MnSb/HfIrSb/Co$_2$MnSb MTJ having different interfaces. Two atomic terminations for HfIrSb along (001) crystal orientation were considered: Ir termination and Hf-Sb termination. The original structures along (001) crystal orientation included Mn-Sb/Ir, Mn-Mn/Ir and Mn-Sb/Hf-Sb , Mn-Mn/Hf-Sb interfacial terminations, which can be divided into two groups: one where interfacial atoms sit on top of Mn or Sb atoms (Top), and the other where interfacial atoms are located in the bridge site between Mn and Sb atoms (Hollow) (shown in Fig. S7 \cite{supply}). Table \ref{tab:1} shows the surface free energy for all the various interfaces considered for the present study. For both terminations, the interface with Ir atoms sitting on the hollow side was found to be energetically favorable. Our DFT calculations were performed by fixing the in-plane lattice constant to the electrode lattice constant, and allowing the hetero-junctions to relax in the z-direction. We prepare the supercell of multilayer containing 17 atomic layers (ML) of Co$_2$MnSb and 13 ML of HfIrSb for the Mn-Sb and
Mn-Mn terminated interfaces, respectively. The selection of the semiconductor layer thickness has been made judiciously, taking into account the smaller band-gap of HfIrSb in comparison to widely used semiconductors like MgO. This  choice enables it to serve as an effective barrier layer while  providing a reasonably high  conductance.  We observe that the off-stoichiometric interface Mn-Mn/Ir had slightly lower surface free energy than the stoichiometric Mn-Sb/Ir interface (Table\ref{tab:1}), and the bond distances at the interface are found to be similar to the Mn-Ir (2.69 \AA) and Ir-Sb (2.70 \AA) bond lengths in the bulk IrMnSb \cite{Tutic_2017}, as shown in Table \ref{tab:1}.\\

\begin{figure*}[!htbp]
	\begin{center}
		\includegraphics[width=1\textwidth] {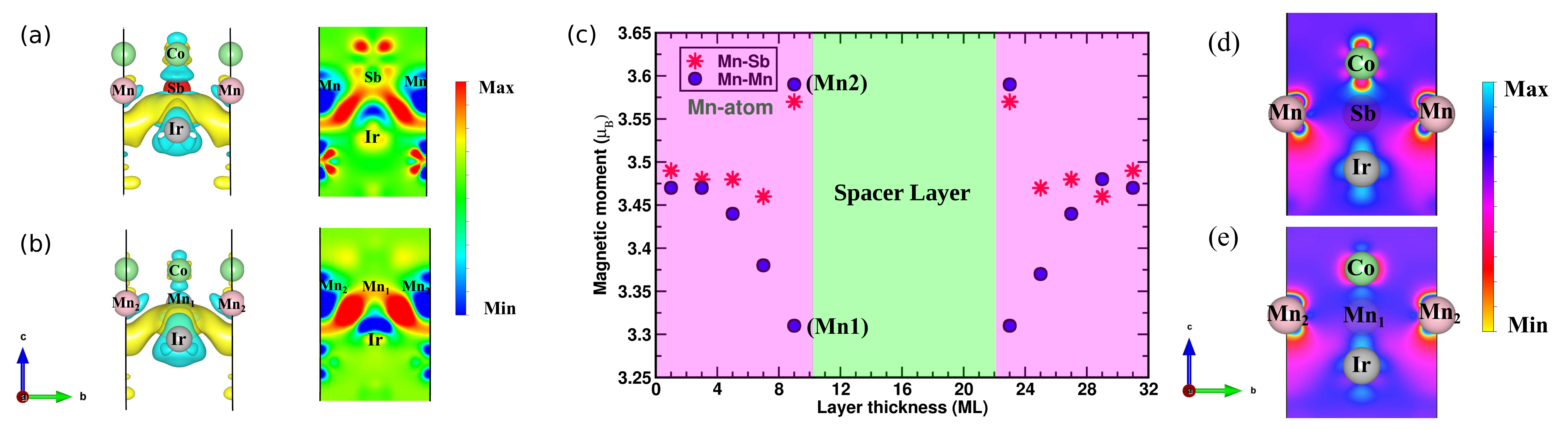}
		\caption{The charge density difference ($\Delta{\rho}$) of the heterostructure is plotted in two forms: a 3D visualization and a 2D visualization in the yz plane. Panels (a) and (b) show the $\Delta{\rho}$ at the Mn-Sb/Ir interface and Mn-Mn/Ir interface, respectively, where blue and yellow colors in the 3D visualization indicate negative and positive $\Delta{\rho}$, respectively. The isosurface value is set to 0.0005 e/A$^3$ for both cases. Panel (c) presents the magnetic moment of the Mn atoms across the Co$_2$MnSb/HfIrSb heterojunction for both the Mn-Sb/Ir and Mn-Mn/Ir interfaces. (d) and (e) depict the magnetization density difference projected in 2D yz plane for Mn-Sb/Ir and Mn-Mn/Ir interface, respectively.}		
		\label{Fig:2}		
	\end{center}
	\end{figure*}

\textbf{Charge Density Difference (CDD):}
In order to understand the chemical bonding at the interface, it is crucial to have a comprehensive understanding of the charge transfer in the system. The transfer of charge ($\Delta{\rho}$) at the interface can be visualized in three dimensions, as well as two dimensions, as shown in Fig. \ref{Fig:2}. The calculation of $\Delta{\rho}$ is carried out by subtracting the spatial charge densities of the electrode and the spacer layers from that of the whole heterostructure, represented by $\rho^{Electrode}$, and $\rho^{Spacer}$, and  $\rho^{MTJ}$, respectively. \\
$\Delta{\rho}$ = $\rho^{MTJ}$ - $\rho^{Electrode}$ -  $\rho^{Spacer}$\\

The value of $\Delta{\rho}$ is positive in the yellow-colored regions and negative in the blue-colored regions. A positive value indicates an accumulation of electronic charge, while a negative value signifies depletion of electronic charge. In Fig. \ref{Fig:2}(a) and (b), it is observed that charge is mostly transferred from the interface Mn and Sb atoms and is accumulated around the interfacial region between Mn-Sb and Ir planes. In Figure S10 \cite{supply}, we present the charge density difference ($\Delta \rho$) at two distinct interfaces (Mn-Sb/Ir and Mn-Mn/Ir), taking into account the influence of spin-orbit coupling (SOC). Remarkably, we find that the charge distribution at the interfaces remains largely unchanged with the inclusion of SOC. However, a notable disparity is observed at the Mn-Mn/Ir interface, revealing an accumulation of charges around the adjacent Hf atom in the subsequent Hf-Sb plane, a phenomenon absent in the absence of SOC.


\textbf{Magnetic Properties :}
In Fig. \ref{Fig:2}(c), the magnetic moments of the interface Mn atoms are shown across the junction for both interfaces. It is observed that the magnetic moments of the interface Mn atoms exhibit a sudden jump. Moreover, for the Mn-Mn/Ir interface, the magnetic moments of the two interface Mn atoms (Mn$_1$ and Mn$_2$) are different. The difference magnetization density plots in Fig. \ref{Fig:2} (d) and (e) indicate that the change in magnetization is mainly localized at the interface Mn atom, while the interface Ir atom acquires a small magnetic moment (-0.1 $\mu_B$). Additionally, the difference magnetization density supports the inequivalence of the magnetization density of the Mn$_1$ and Mn$_2$ atoms for the Mn-Mn/Ir interface.

It should be emphasized that the difference magnetization density around the interface Mn atoms arises primarily due to the localized d$_{yz, xz}$ orbitals. Furthermore, the small magnetic moment of the Ir atom at the interface induced by the proximity effect of the magnetic layer arises mainly from the out-of-plane d$_{z^2}$ orbitals, as shown in Fig. \ref{Fig:2} (d), (e).  \\

\begin{table*}[!htbp]
		\footnotesize
		\begin{center}
			
			\caption{The calculated interface free energies of various optimized Co$_2$MnSb/HfIrSb interface heterostructures are presented, along with the corresponding bond lengths between interfacial atoms and the observed spin-polarization (SP) at the interface of the heterostructure showing lowest surface free energy for each termination. }
			\label{tab:1}
			\begin{tabular}{c c c c c c c c c  }
				\hline
				\hline

				& Surface termination& Atomic position  &Interfaced &Surface free energy (in eV/\AA$^2$) &Bond
Type& Bond-length (in \AA)& Interface SP (in $\%$)&
				\\ 
    \hline\hline
&Mn-Sb&Top&MnSb|Ir&-6.8106&&&&
\\
&&&MnSb|HfSb&-6.8122&&&&
\\
&&Hollow&\textbf{MnSb|Ir}&-6.9206&Mn-Ir, Sb-Ir&2.69, 2.70&68&
\\
&&&MnSb|HfSb&-6.8482&&&&	
\\&Mn-Mn&Top&MnMn|Ir&-6.9141&&&&
\\
&&&MnMn|HfSb&-6.8312&&&&
\\
&&Hollow&\textbf{MnMn|Ir}&-6.9825&Mn1-Ir, Mn2-Ir&2.60, 2.57&48&
\\
&&&MnMn|HfSb&-6.7456&&&&	
\\
\hline\hline
					
			\end{tabular} 
		\end{center}
	\end{table*}

\begin{figure}
	\begin{center}
		\includegraphics[width=60mm,scale=0.5] {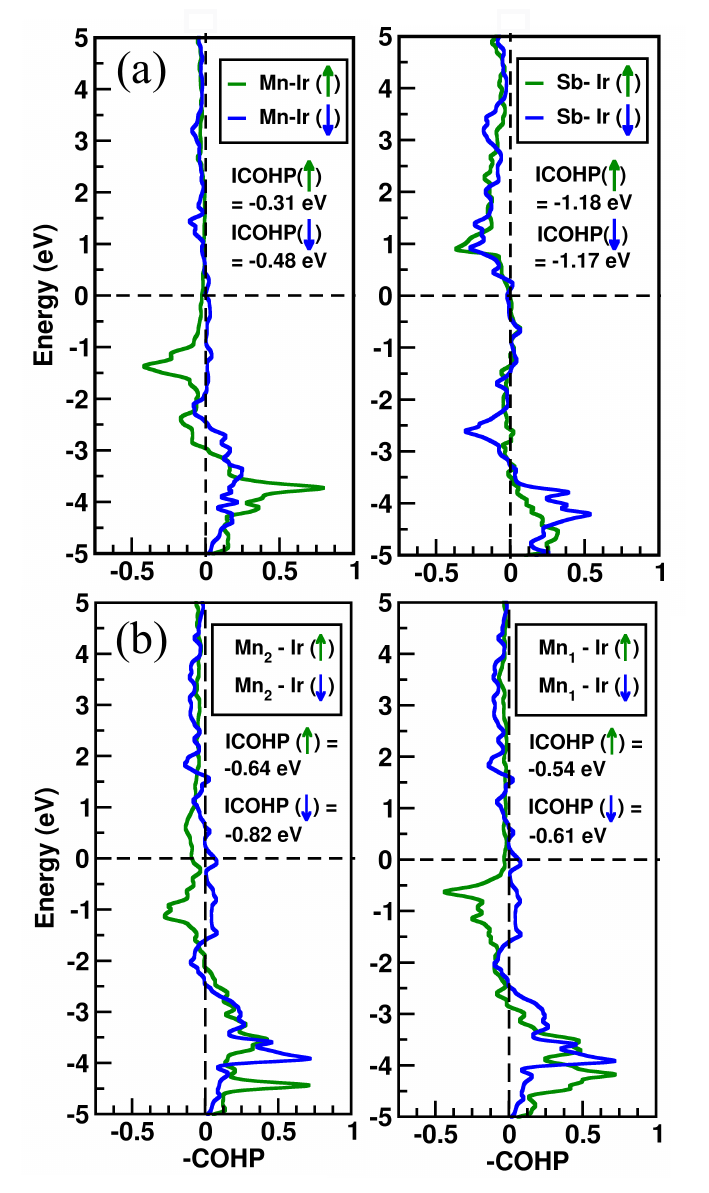}
		\caption{COHP analysis of the bonds between the interface atoms for Co$_2$MnSb/HfIrSb heterojunction: (a) Mn-Sb/Ir ; (b) Mn-Mn/Ir interface, respectively. E$_F$ has been set to 0 eV. The positive (negative) values at the x-axis indicates bonding (anti-bonding) feature. The sign $\uparrow$($\downarrow$) indicate majority (minority) spin contributing to the bonding. }		
		\label{Fig:3}		
	\end{center}
\end{figure}	

\textbf{Bader Charge and COHP Bond Analysis:}
The net charge transfer at Co$_2$MnSb/HfIrSb interfaces as calculated from Bader charge analysis \cite{HENKELMAN2006354} is shown in Fig.S9 \cite{supply}. For both the interfaces we observe that the Ir atom at the interface loses significant amount of charge (-0.27e) compared to the interface Mn (-0.14e) and Sb (-0.07e) atom, that get accumulated in the interface region. Similar trend can also be observed from our Mulliken and Loewdin
charge analysis ( Table S4. \cite{supply}) as obtained from LOBSTER package\cite{Dronskowski1993}. Additionally, our orbital-decomposed charge density analysis, facilitated by the LOBSTER package \cite{Dronskowski1993}, highlights that out-of-plane orbitals ($s$ and $d_{z^2}$ for Ir and $p_z$ for Sb) witness the most significant charge losses, whereas for Mn, it is the in-plane ($d_{x^2-y^2},~d_{xz}$) orbital that experiences substantial charge loss. These findings corroborates well with our charge density and magnetization density difference analysis presented in Fig.\ref{Fig:2}.\\

To gain deeper insights into the interface bonding, we conducted the Crystal Orbital Hamilton Population (COHP) analysis between Mn-Ir and Sb-Ir atom pairs, as depicted in Fig.\ref{Fig:3}. Additionally, in Fig.\ref{Fig:3}, we present integrated COHP values (ICOHP) between these atom pairs, serving as an indicator of bond strength by integrating the COHP values up to the Fermi energy (E$_F$). For the Mn-Sb/Ir interface (Fig.\ref{Fig:3} (a)), the COHP analysis between Mn-Ir atom reveals presence of anti-bonding states near E$_F$ in the majority spin channels, which lowers the bonding interaction. However, the situation is different for the minority-spin channel, where the whole valance band (VB) has bonding interaction. Similar observations have also been made in some previous bonding analysis involving $3d$ transition metal atoms \cite{Nelson2017,Richard,BHATTACHARYA2021}. Because, of the exchange hole, the majority spin orbitals are more spatially contracted than the minority spin orbitals and contributes less to the bonding interactions \cite{Nelson2017}. The COHP bonding analysis between the Ir-Sb pair mostly revealed bonding interaction in the valance band for both the spin channel, apart from the presence of small anti-bonding interaction in the minority spin-channel in the VB away from E$_F$, which is compensated by the strong bonding character deep into the energy. The ICOHP values in Fig.\ref{Fig:3} indicates the Sb-Ir (ICOHP:-2.35 eV) bonding is significantly stronger than the Mn-Ir (ICOHP:-0.79 eV ) bonding at the interface. \\
In the case of the Mn-Mn/Ir interface, the COHP analysis uncovers anti-bonding states in the majority spin channels near E$_F$ between the both, Mn$_1$-Ir and Mn$_2$-Ir bonds, contributing to even weaker bonding compared to the Mn-Sb/Ir interface. In contrast, the minority-spin contributes exclusively to the bonding interaction below E$_F$ (Fig.\ref{Fig:3} (b)). \\
The weaker covalent bonding at the Mn-Mn/Ir interface compared to the Mn-Sb/Ir interface, as indicated by our COHP bonding analysis is further supported by the Bader charge analysis, which suggests relatively less charge transfer at the Mn-Mn/Ir interface compared to the Mn-Sb/Ir Interface. 

\begin{figure}
	\begin{center}
		\includegraphics[width=.5\textwidth] {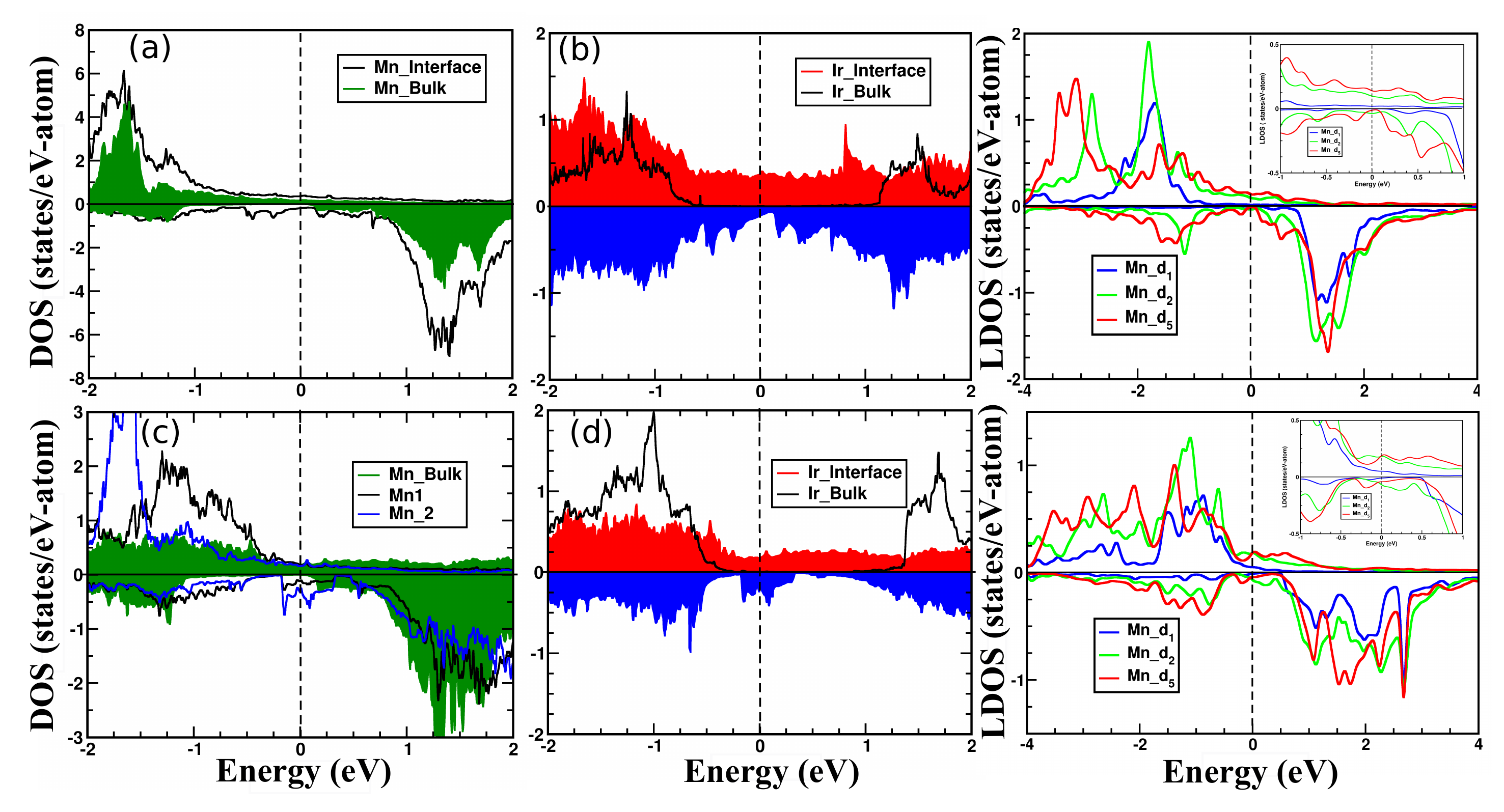}
		\caption{The local density of states (LDOS) for the interfacial Mn, Sb, and Ir atoms are shown in panels (a), (b), and (d), (e), respectively, for the heterojunctions with Mn-Sb/Ir and Mn-Mn/Ir interfaces. The LDOSs in the bulk region are also presented as a reference using filled curves in each figure. Panels (c) and (f) display the orbital projected DOS of the interfacial Mn atom for the heterojunctions with Mn-Sb/Ir and Mn-Mn/Ir interfaces, respectively, with d$_1$, d$_2$, d$_3$ denoting d$_{z^2}$, (d$_{xy}$, d$_{x^2 -y^2}$), and (d$_{yz}$ d$_{xz}$), respectively.}		
				\label{Fig:4}		
	\end{center}
\end{figure}	
\textbf {Electronic Density of States $\&$ Band Structure:}
To investigate the influence of interfacial interactions on electronic behavior, we analyze the projected DOS of the interface atoms, as depicted in Fig. \ref{Fig:4}. The atom projected DOS of the interfacial Mn atoms for both the Mn-Sb/Ir and Mn-Mn/Ir interfaces are shown in Fig. \ref{Fig:4} (a) and (d), respectively. Our analysis reveals that the half-metallic character of the bulk is disrupted for both interfaces. However, the Mn-Sb/Ir interface still exhibits a high degree of spin polarization (approximately 68$\%$). The DOS of the different interface Mn atoms for the Mn-Mn/Ir interface significantly differ from each other. Specifically, the Mn$_1$ atom displays high majority spin density at around -1 eV, while the Mn$_2$ atom exhibits a slightly shifted density of states, which is towards the higher binding energy side (approximately -1.8 eV). This results in a higher exchange splitting energy and thus larger magnetic moment for the Mn$_2$ atom is observed.  \\

We show various $d$-orbital contributions of the Mn atom for the Mn-Sb/Ir and Mn-Mn/Ir interfaces in Fig. \ref{Fig:4}(c) and (f), respectively. Here, d$_1$, d$_2$, d$_3$ correspond to the d$_{z^2}$, (d$_{xy}$, d$_{x^2-y^2}$), and (d$_{yz}$, d$_{xz}$) orbitals of the interface Mn atoms, respectively. Though the formation of the heterojunction breaks the periodicity along the z-direction, the d$_1$ orbital retains the half-metallic character of the bulk. However, the d$_2$ and d$_5$ orbitals are primarily responsible for the destruction of the half-metallicity at the interface. There is a significant difference between the majority spin DOS at the E$_F$ for the two interfaces. For the Mn-Sb/Ir interface, there is a dominant contribution from the d$_5$ orbitals (Fig. \ref{Fig:4}(c)), while for the Mn-Mn/Ir interface, both d$_2$ and d$_5$ orbitals have a significant contribution. Since the orbital characters of the traveling electrons play a crucial role in transmission, one would expect differences in the spin-transmission behavior between these two interfaces, which are discussed in the subsequent section. \\

The Ir atom at the interface becomes metallic for both interfaces, while the semiconducting behaviors persist in the bulk region (Fig. \ref{Fig:4} (b), (e)). Consequently, there is a small gain in magnetic moments (approximately -0.1 $\mu_B$) for the interface Ir atoms for both interfaces. \\

Figure. S11 \cite{supply} shows a comparison of total DOS of the interface atoms for each interfaces of Co$_2$MnSb/HfIrSb heterojunction, with considering the effect of SOC and without SOC. As depicted in Figure. S11 \cite{supply}, the DOS around the E$_F$ gets hardly affected by SOC. Therefor we do not include the effect of SOC in the following part of our discussion.
\begin{figure}
	\begin{center}
		\includegraphics[width=.5\textwidth] {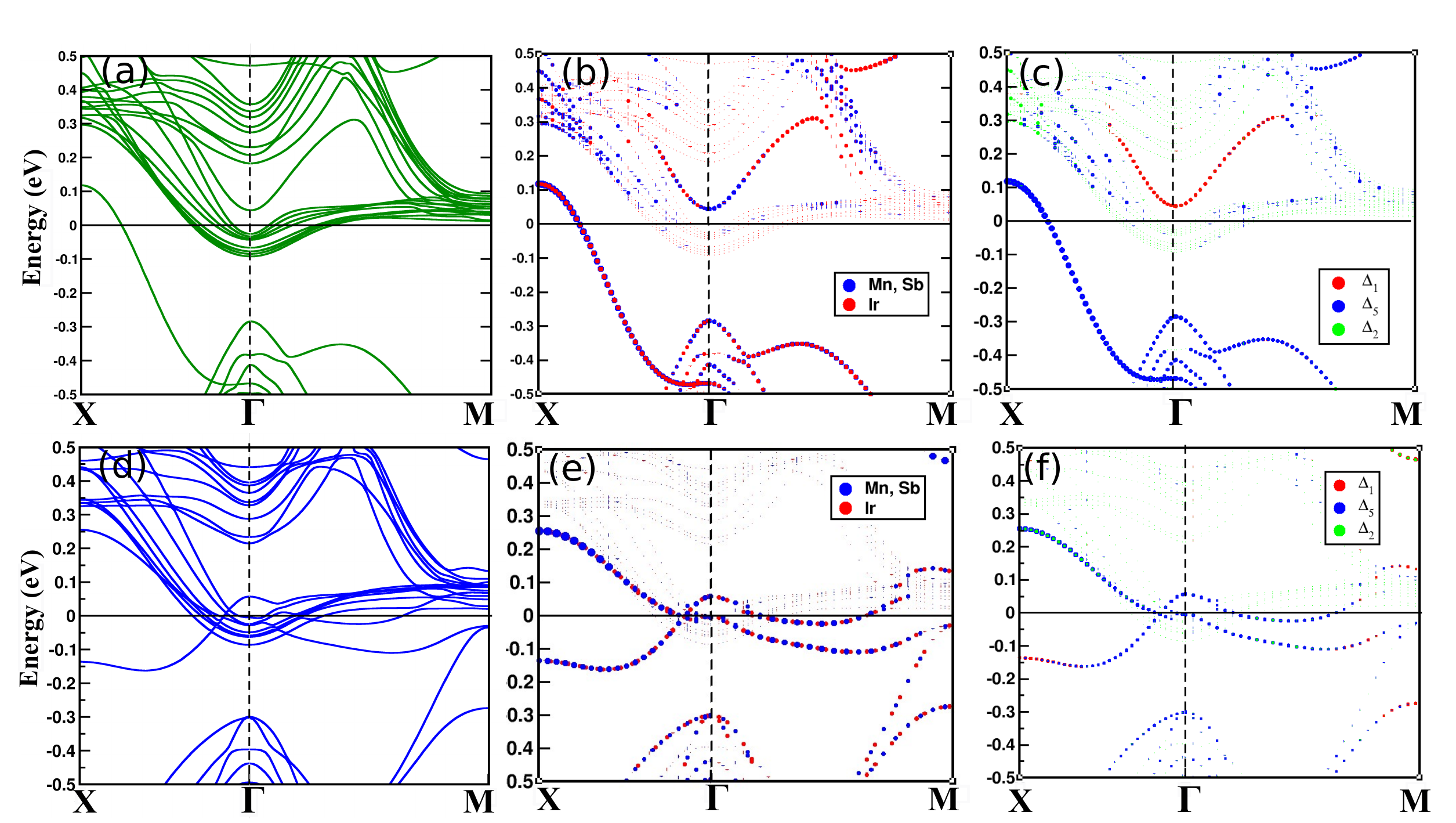}
		\caption{(a) Minority spin band structure for the  considered heterojunctions with Mn-Sb/Ir (Mn-Mn/Ir) interface on the top (bottom) panel along the X-$\Gamma$-M high symmetry directions of the 2D Brillouin zone; (b), (c) represent the atomic and orbital contribution of the inter-facial atoms on the minority spin band structures, respectively. Here $\Delta_1$ $\Delta_2$ and $\Delta_5$ represent ( s, p$_z$, d$_{z^2}$) , (d$_{xy}$, d$_{x^2 -y^2}$), and (p$_x$, p$_y$, d$_{yz}$ d$_{xz}$) orbitals respectively. }		
		\label{Fig:5}		
	\end{center}
\end{figure}

To achieve a high TMR ratio, it is essential to minimize the current passing through the barrier when the magnetization of the electrodes is anti-parallel. When perfect half-metallic electrodes are used, spin flipping and tunneling to or from an interfacial state can produce current in an anti-parallel configuration. The tunneling probability of carriers in various bands can vary significantly depending on their band symmetry, as demonstrated in the literature.\cite{Peter,Miura} Electrons in states with $\Delta_1$ orbital symmetry exhibit weak decay within the barrier material, whereas the transmission of electrons in other symmetry states is exponentially suppressed. Our prior research \cite{BHATTACHARYA_2023} illustrated the presence of $\Delta_1$ symmetric bands in the majority spin channel for Co$_2$MnSb. This suggests that it is easy for majority spin electrons to tunnel through the barrier in a parallel spin configuration, which is a prerequisite for achieving a high TMR.
 
Nonetheless, it is equally important to reduce the tunneling rate into minority interface states to suppress the current for anti-parallel magnetization. We present the band structure of minority spin states for the heterojunction with Mn-Sb/Ir and Mn-Mn/Ir interfaces in Fig. \ref{Fig:5}. In Fig. \ref{Fig:5}(b) and (e), we demonstrate the contribution of interface atoms (i.e., Mn, Sb, and Ir) to the minority spin bands.
The orbital projected band structures of the interface atoms in Fig. \ref{Fig:5}(c)  imply that the minority spin conduction bands for the Mn-Sb/Ir interface have a dominant $\Delta_1$ character, suggesting a larger transmission for the minority states. Conversely, for the Mn-Mn/Ir interface (Fig. \ref{Fig:5}(f)), mostly in-plane d orbitals dominate the minority spin states near E$_F$, leading to poor coupling with the $\Delta_1$ type bands of the HfIrSb spacer material. 

\subsubsection{TMR Ratio and the Effect of Electric Field}
\begin{figure}
		\begin{center}
			\includegraphics[width=.5\textwidth] {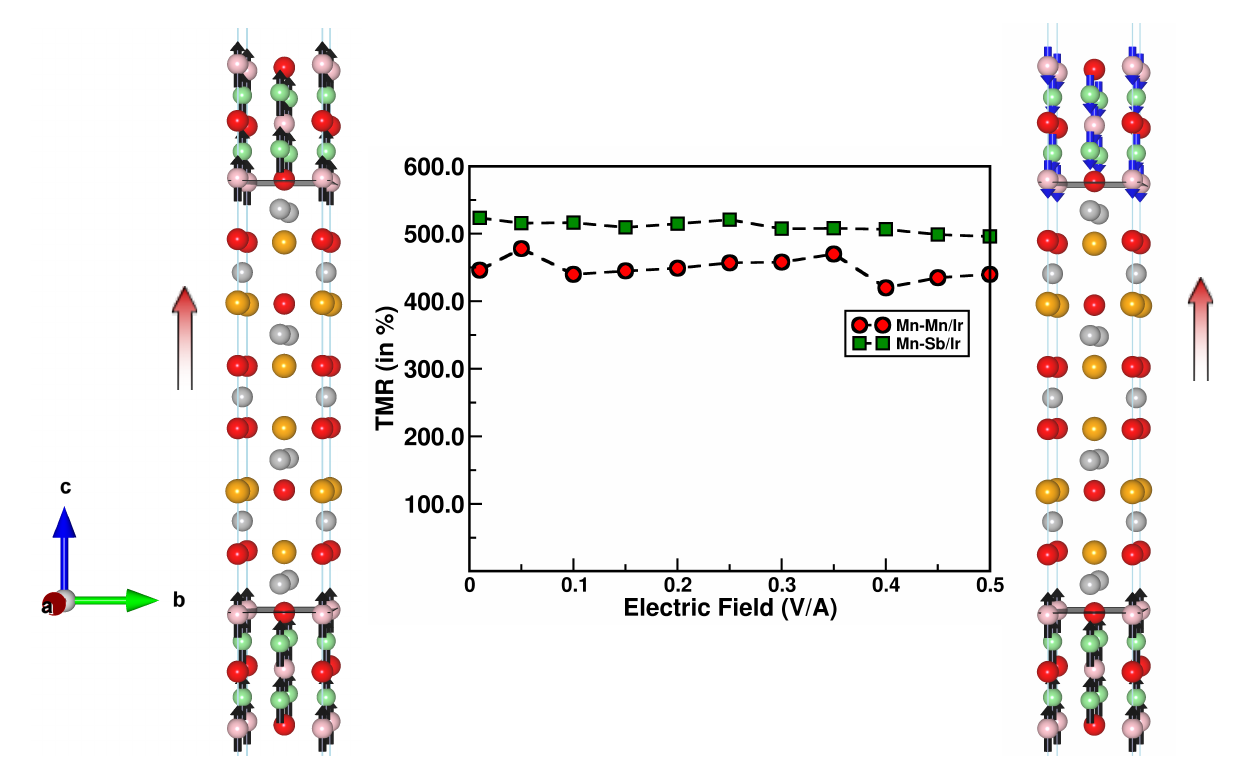}
			\caption{The change in tunnel magnetoresistance (TMR) ratio of the heterojunction system as a function of an applied external electric field is depicted. The schematic representation of the heterojunction includes a top and bottom magnetic layer. The left (right) panel illustrates the ferromagnetic (anti-ferromagnetic) alignment of spins within the magnetic layer. The orientation of magnetization is denoted by black and blue arrows, while the red arrow indicates the direction of the electric field. }		
			\label{Fig:6}		
		\end{center}
	\end{figure}

Next, we aim to investigate the influence of an external electric field on the TMR ratio of the heterojunction. Our objective is to explore the potential for achieving electrical control of magnetic tunnel junctions (MTJs). The TMR ratio has been calculated according to the Eq. [1] based on the standard model given by Julliere.\cite{JULLIERE} As depicted in Fig. \ref{Fig:6}, we observe that the TMR values for both the Mn-Sb/Ir and Mn-Mn/Ir interfaces exhibit minimal sensitivity to the external electric field. Significant TMR values are observed for both interfaces. However, we do not observe any magnetoelectric coupling, as has been reported in some other MTJs.\cite{Sun_2020,Zhang2022,Bai_2014} Additionally, the application of an external electric field does not significantly affect the magnetic moments at the interfaces. Instead, we observe a proximity effect where the interface Ir atoms acquire small magnetic moments (approximately 0.10 $\mu_B$), and the direction of these moments depends on the magnetization direction of the adjacent Mn atoms. In Figure. S12 \cite{supply} we have further shown the charge density difference ($\Delta\rho$) of Co$_2$MnSb/HfIrSb heterojunction, featuring the Mn-Sb/Ir interface, in the presence and absence of an external electric field. This clearly depicts that the charge distribution at the interface remains unaffected under the application of external electric field. We only observe difference in charges at those layers, which are exposed to the vacuum. The absence of electric field control over the magnetic properties in the all-Heusler alloy junctions can be attributed to the weaker covalent bonding between the interface atoms, as suggested by the negligible interface buckling observed. This weaker bonding stands in contrast to other Heusler alloy and oxide-based magnetic tunneling junctions, which exhibit greater electric field control.\cite{Bai_2014}

\subsubsection{Spin-Transport Properties}	
\begin{figure*}[!htbp]
	\begin{center}
		\includegraphics[width=1\textwidth] {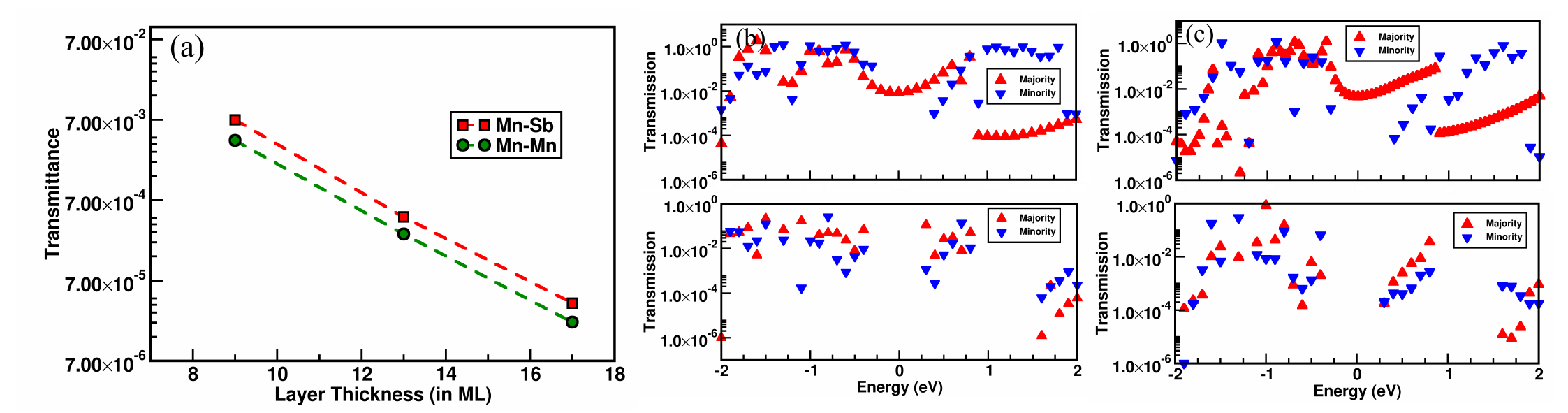}
		\caption{(a) Transmittance in parallel magnetization case for Co$_2$MnSb/HfIrSb/Co$_2$MnSb junction with varying thicknesses of HfIrSb layers; (b), (c) Energy-dependent spin-resolved transmission coefficients for the same with 13 ML of HfIrSb layer with Mn-Sb/Ir and Mn-Mn/Ir interface, top (bottom) panel is for parallel (anti-parallel) spin configuration of the electrodes, respectively. All the graphs are plotted in log scale. In panels (b), (c) E$_F$ has been set to zero.}		
		\label{Fig:7}		
	\end{center}
	\end{figure*}

Our focus now turns to examining the spin-resolved transport properties of the heterojunction . We present the  results for the parallel (P) and anti-parallel (AP) spin alignments in Fig. \ref{Fig:7}. Fig.\ref{Fig:7} (a) shows the majority spin transmittance (in log scale) as a function of spacer layer thickness for both interfaces. The exponential decay of the transmittance with spacer layer thickness confirms the tunneling behavior of the heterojunction \cite{Roy_2022,Butler_2008}. Figures \ref{Fig:7}(b) and (c) illustrate the energy-dependent spin transmission for the heterojunction with Mn-Sb/Ir and Mn-Mn/Ir interfaces, in P and AP spin configuration of the electrodes, respectively. Upon examining the majority spin transmission for both the heterojunctions, we observe that two distinct features emerge. Firstly, in the energy range from $\approx$ -0.61 eV to 0.15 eV, the transmission decays, suggesting the tunneling of electrons with energies lower than the barrier. Secondly, a sudden drop in transmission occurs around 1 eV, since the $\Delta_1$ band of the Co$_2$MnSb electrode, which is the primary contributor to electron tunneling, extends up to just below 1 eV (see Fig. S1(d)\cite{supply}). Beyond this energy, it is mostly the bands with $\Delta_5$ symmetry that contribute to the tunneling. For both spin channels and magnetic configurations in Fig.\ref{Fig:7} (b), (c) E$_F$ of the Co$_2$MnSb/HfIrSb/Co$_2$MnSb junction lies close to the valence band of the bandgap of HfIrSb. Due the HM nature of the electrode, we do not observe any transmission at  E$_F$ for the minority spin channel as well as for majority and minority spin channels for AP configuration of the electrodes . For the AP configuration, as expected, the transmission
coefficients are nearly  (not exactly, because the inversion symmetry is broken) spin-degenerate. The transmission behavior that we have described indicates a significant level of spin-filtering in the junction.
 
\begin{figure}
	\begin{center}
		\includegraphics[width=.5\textwidth] {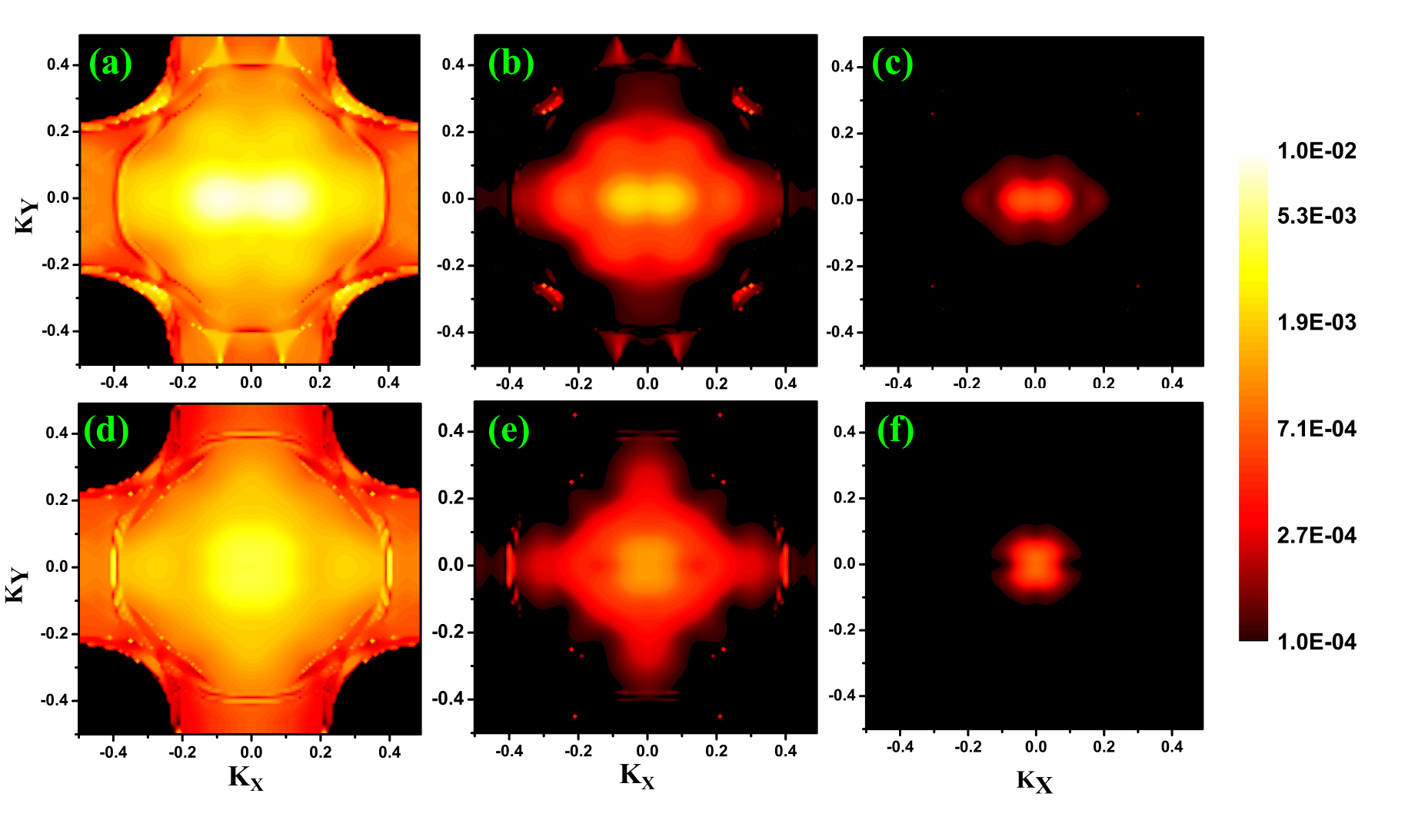}
		\caption{Dependence of the majority spin transmission (values plotted in log scale) over the 2D Brillouin zone at E$_F$ in parallel spin configuration with Mn-Sb/Ir (Mn-Mn/Ir) interface in top (bottom) panel. (a), (b), (c) ((d), (e), (f) ) are for 9ML, 13Ml, 17ML of HfIrSb layer, respectively. }		
		\label{Fig:8}		
	\end{center}
	\end{figure}	

\begin{figure*}[!htbp]
	\begin{center}
		\includegraphics[width=.75\textwidth] {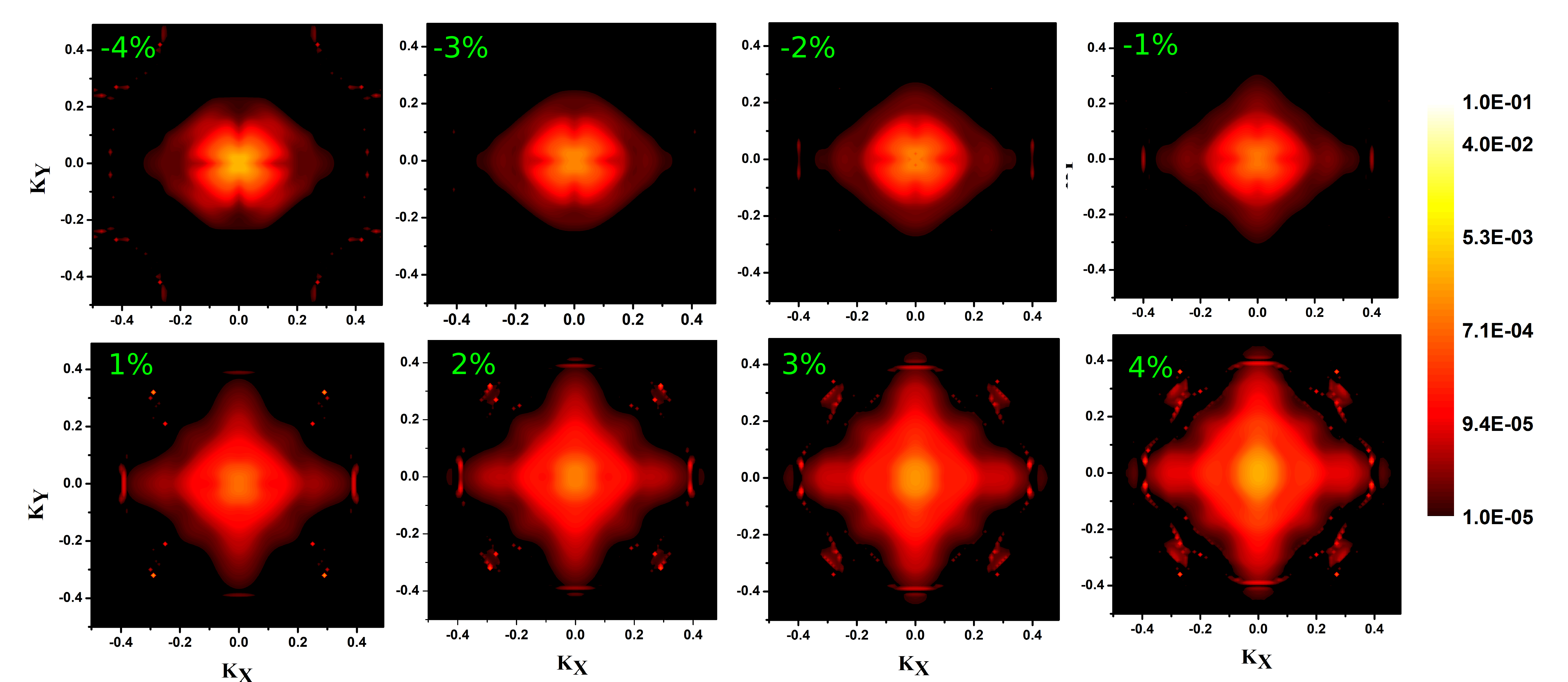}
		\caption{Dependence of majority spin transmission (values plotted in log scale) over the 2D Brillouin zone at E$_F$ in parallel spin configuration with Mn-Mn/Ir interface corresponding to a applied compressive (-4$\%$ to -1$\%$ ) and tensile (+1$\%$ to +4$\%$) bi-axial strain. The spacer layer thickness has been chosen to 13 ML.  }		
		\label{Fig:9}	  
	\end{center}
	\end{figure*}

\begin{figure*}[!htbp]
	\begin{center}
		\includegraphics[width=.6\textwidth] {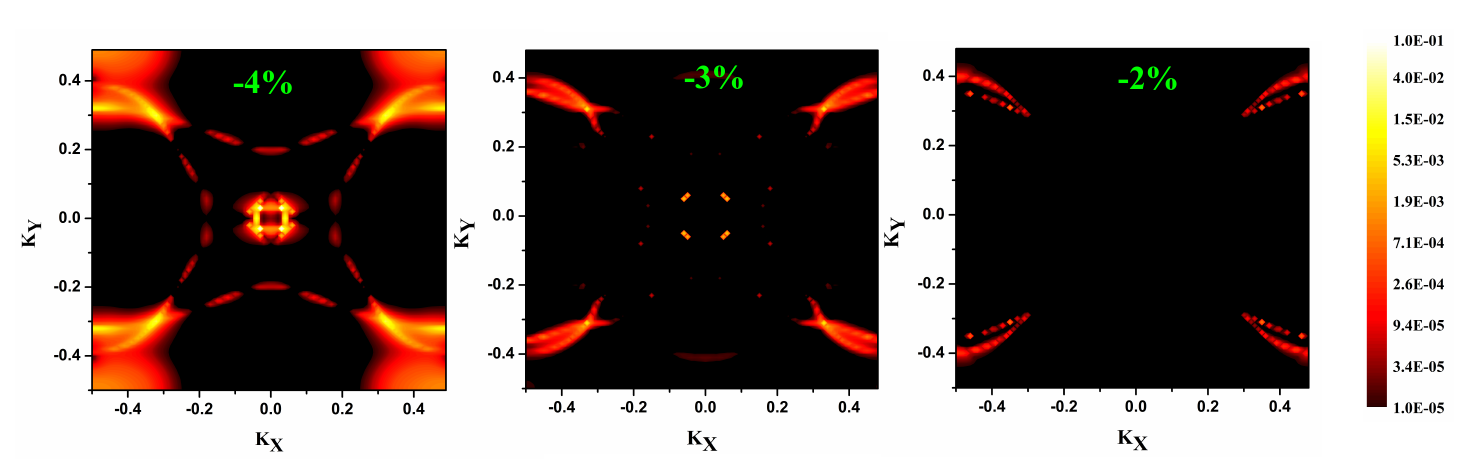}
		\caption{Dependence of minority spin transmission (values given in log scale) over the 2D Brillouin zone at E$_F$ in parallel spin configuration with Mn-Mn/Ir interface corresponding to a applied (compressive)  bi-axial strain of -4$\%$ to -2$\%$, beyond that no  transmission is observed for the minority spin states . }		
		\label{Fig:10}	
	\end{center}
\end{figure*}

 \begin{figure*}[!htbp]
	\begin{center}
		\includegraphics[width=.75\textwidth] {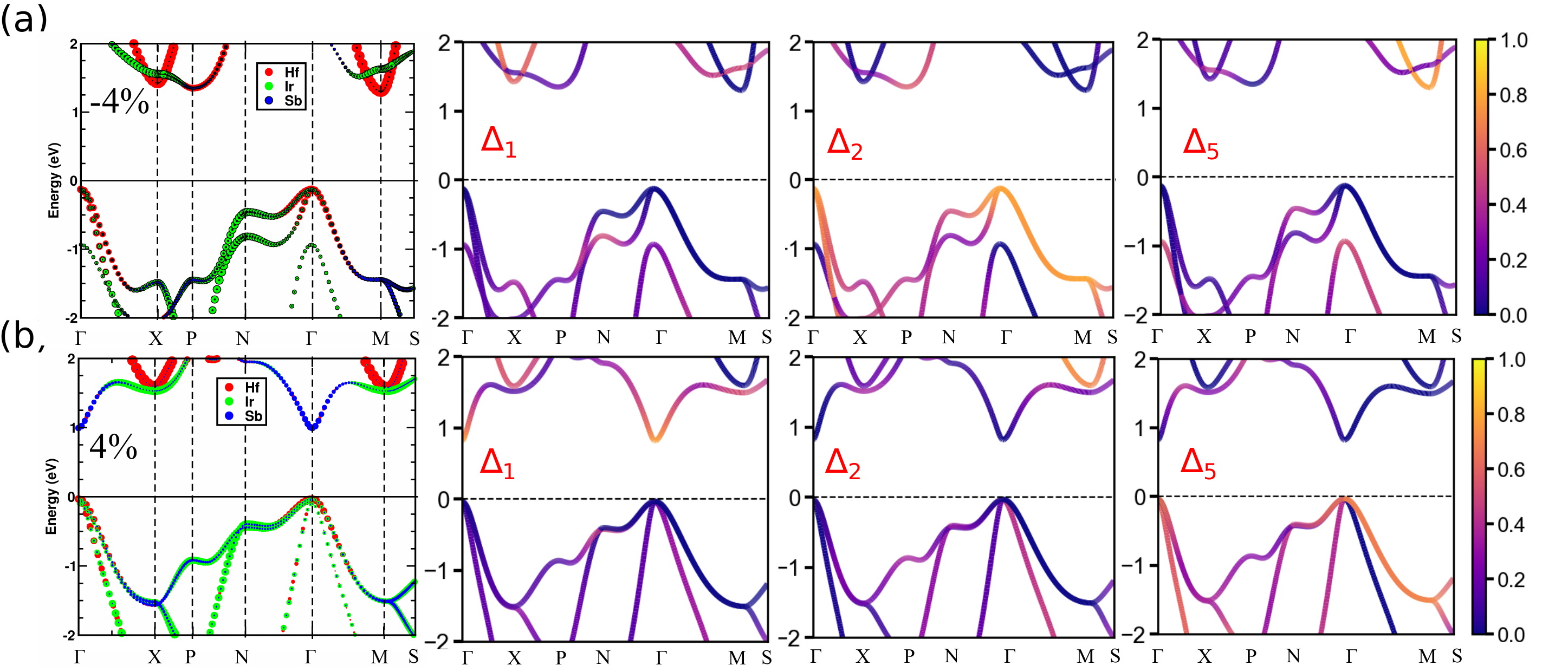}
		\caption{The atom and orbital projected band structure of bulk HfIrSb under strain is investigated, specifically for (a) -4$\%$ (compressive) strain and (b) +4$\%$ (tensile) strain. In this analysis, the symbols $\Delta_1$, $\Delta_2$, and $\Delta_5$ represent the following orbital compositions, respectively: (s, p$_z$, d$_{z^2}$), (d$_{xy}$, d$_{x^2 -y^2}$), and (p$_x$, p$_y$, d$_{yz}$, d$_{xz}$). }		
		\label{Fig:11}	
  \end{center}
	\end{figure*}

In Fig. \ref{Fig:8}, we present the k-dependent majority spin transmission in the 2D-Brillouin Zone at E$_F$ for two heterojunctions with varying spacer layer thicknesses. Our analysis of the k-dependent transmission reveals that the transmission predominantly occurs around the lowest decay point, which is located at $\Gamma$ for both the interfaces. However, there are notable differences in the transmission profiles for the two interfaces. For the Mn-Sb/Ir interface, the transmission profile exhibits 2-fold rotational symmetry, whereas for the Mn-Mn/Ir interface, it shows a combination of 4-fold and 2-fold rotational symmetry. This distinction suggests that the majority spin carriers for these two interfaces have different orbital symmetries. We have shown the orbital projected DOS for the majority spin states at E$_F$ in Fig. \ref{Fig:3}(c) and (f), which indicates that d$_{xz,yz}$ orbitals have a significant contribution for the Mn-Sb/Ir interface, while for Mn-Mn/Ir, d$_{xz,yz}$ as well as d$_{x^2-y^2,xy}$ orbitals play an important role. This observation explains the difference in the transmission profiles for the two cases, where the heterostructure with Mn-Sb/Ir interface exhibits majority spin transmission dominated by electrons with $\Delta_5$ orbital symmetry, while for Mn-Mn/Ir, the $\Delta_2$ and $\Delta_5$ states dominate. Previous studies have established that electrons with $\Delta_2$ orbital symmetric states decay faster than those with $\Delta_5$ states inside the barrier region. This is further supported by the variation of the absolute square of the scattering wavefunction at E$_F$ as a function of heterojunction layer thickness (as depicted in Figure. S13) \cite{supply}. This shows for the Mn-Mn/Ir interface the scattering states inside the barrier decays faster than of Mn-Sb/Ir interface.  Consequently, we observe that the majority spin transmittance with Mn-Mn/Ir interface is smaller than that with the Mn-Sb/Ir interface (Fig. \ref{Fig:7} (a)). Additionally, we notice that with increasing barrier thickness, the transmissions of tunneling electrons with finite k$_{||}$ values are highly suppressed due to their faster decay compared to those with k$_{||}$ =0 and the transmission is mostly centered around the $\Gamma$ (Fig. \ref{Fig:8} (c) and (f)). We, further observe that with increasing spacer layer thickness, the transmission due to $\Delta_2$ orbital symmetric bands at the Mn-Mn/Ir interface is getting suppressed and transmission due to the $\Delta_5$ orbitals symmetric bands sustains (Fig.\ref{Fig:8} (e)).  \\

\subsubsection{Effect of Strain Engineering on spin-transport properties}
Strain engineering offers a promising approach to control the electronic and magnetic properties of a material, which is essential for improving the performance of spintronic devices. Several studies in the literature demonstrate that modifying the TMR and magneto-crystalline anisotropy of MTJs using strain engineering is feasible.\cite{Loong2014,roschewsky2018perpendicular} It is also reported that strain-mediated switching mechanisms can reduce the energy barrier between parallel and antiparallel states.\cite{Noh_19,Biswas2017} However, previous research on the effect of strain on Heusler alloy-based MTJs mainly focuses on half-metallicity, spin polarization, and magnetic moment. In this study, we also aim to investigate the impact of strain engineering on the spin-transmission properties of the heterojunction. This is especially relevant because the selected material in the present study exhibits approximately +5$\%$ lattice mismatch, indicating that the investigation on the effect of strain on transport properties might be  crucial.

To achieve strain engineering, we chose a heterojunction with an Mn-Mn/Ir interface that exhibits lower surface free energy
(Table \ref{tab:1}). Further analysis of the electronic properties of this heterostructure at the interface reveals competing contributions from $\Delta_2$ and $\Delta_5$ states in the majority spin states, indicating that the interface might show sensitivity to the bi-axial strain. Therefore, we investigate the majority and minority spin transmission of the heterojunction with Mn-Mn/Ir interface over the 2D Brillouin zone under  bi-axial strain ranging from -4$\%$ to 4$\%$ as shown in Fig. \ref{Fig:9} and \ref{Fig:10}. 

The bi-axial strain was applied by fixing the in-plane lattice constant and allowing the volume of the heterojunction to relax. Our results show that bi-axial strains have a significant effect on the majority spin-transmission property, with a change in the transmission profile from 4-fold rotational symmetry to 2-fold rotational symmetry as bi-axial strain changes from compressive to tensile (Fig. \ref{Fig:9}). This implies that the $\Delta_2$ bands dominate the majority spin transmission under compressive strain, while $\Delta_5$ bands dominate under tensile strain. Additionally, we observe a three-fold increase in the conductance of the majority spin channel from 2.26 $\times$ 10$^{-5}$ $(\frac{e^2}{h})$ to 6.94 $\times$ 10$^{-5}$ $(\frac{e^2}{h})$ as bi-axial strain changed from -4$\%$ to 4$\%$, which is consistent with the weaker decay rate of $\Delta_5$ bands inside the barrier compared to the $\Delta_2$ bands.\\

Under compressive bi-axial strain, the HM property of the electrode is disrupted, resulting in the observation of transmission due to minority spin states (Fig. \ref{Fig:10}). Specifically, for 4$\%$ compressive strain, transmission spots are detected around the $\Gamma$ point and at the corners of the two-dimensional Brillouin zone (Fig. \ref{Fig:10}). However, these transmission spots diminish in size as the compressive strain decreases, and at 1$\%$ compressive strain, the minority transmission becomes completely absent. Notably, under compressive strain, a considerable amount of minority spin transmittance is observed, ranging from approximately $10^{-5}$ to $10^{-6}$. This can be attributed to the presence of $\Delta_1$ symmetric bands in the minority spin states of Co$_2$MnSb under compressive strain (refer to Fig. S3 in the Supplementary Information\cite{supply}). Consequently, this will lead to further decrement of the TMR ratio of the MTJ under compressive strain due to these effects. \\

In order to investigate the underlying mechanism of such orbital sensitive majority transmission, we have conducted a detailed analysis of the electronic properties of the bulk electrode and spacer material. Our previous discussion has revealed that bi-axial strain induces changes in the spin polarization of the electrode, resulting in the loss of its HM property under compressive bi-axial strain (as shown in Fig. S2\cite{supply}). However, we have not observed any significant changes in the orbital character of the bands under the entire range of applied strain along the transport direction ($\Gamma$ to X (Z), as depicted in Figures S3, S4\cite{supply}. Somewhat rigid shift of the bands can only be seen. 

Additionally, we have demonstrated the effect of bi-axial strain on the electronic properties of the bulk spacer material HfIrSb, as illustrated in Fig. \ref{Fig:11}. Specifically, under -4$\%$  bi-axial strain, HfIrSb becomes an indirect bandgap semiconductor, with an increase in the bandgap of 1.25 eV compared to the unstained structure, and the degeneracy of the valence bands at the $\Gamma$ point is also lifted. Furthermore, the atom-projected band structure analysis suggests that both the VBM at $\Gamma$ and CBM at the M point have a dominant contribution from the Hf atoms (as shown in Fig. \ref{Fig:11}(a)), which is unlike the case of the unstained structure (as shown in Fig. \ref{Fig:1}). These changes are also reflected in the orbital-projected band structure (as presented in Fig. \ref{Fig:11} (a)), where we observe that the VBM and CBM have $\Delta_2$ and $\Delta_5$ orbital characters, respectively. 

On the other hand, under +4$\%$  bi-axial strain, we observe that the CBM is mostly Sb atom derived and the VBM has a contribution from both Hf and Ir atoms (as illustrated in Fig. \ref{Fig:11}(b)). The orbital-projected band structure shows,  the VBM is mostly dominated by the $\Delta_5$ orbitals. The comprehensive differences in the electronic properties of the spacer layer under compressive and tensile bi-axial strain lead to the orbital sensitivity of the majority spin transmission. 

Finally, we analyze the results corresponding to the $d$ orbitals of the interfacial Mn atoms of the heterojunction with Mn-Mn/Ir interface under -4$\%$ (compressive) and +4$\%$ (tensile) strain (Fig. S14\cite{supply}). Our results clearly indicate that under -4$\%$  strain, the $d_{x^2-y^2, xy}$ orbitals of the Mn atoms have a significant contribution to the majority spin states at the Fermi level, whereas it is the $d_{xz,yz}$ orbitals of Mn atom that contributes more for +4$\%$  strain. These observations collectively explain the orbital sensitivity of the majority spin transmission under strain.\\

 Our comprehensive observations indicate that the transmission properties in this heterojunction can be effectively tuned by strain. A significant discovery from our calculations is that tensile strain enhances the transmission of the majority spin states, while it considerably hinders the transmission of minority spin states. Considering the lattice mismatch of +5$\%$ between our electrode and spacer materials, this phenomenon can be utilized to achieve even higher TMR ratios under tensile strain. In essence, the application of tensile strain can further optimize the performance of the heterojunction and enhance the TMR ratio.

\subsection{Conclusion}	
Using first-principles density functional theory calculations, we have explored electronic and transport properties of a Co$_2$MnSb/HfIrSb/Co$_2$MnSb all-Heusler magnetic tunneling junction. We demonstrate that the Mn-Sb and Mn-Mn terminated surfaces of Co$_2$MnSb along the (001) direction preserve the half-metallic properties of the bulk. From the surface free energies, we propose that heterojunctions of half-metallic and ferromagnetic alloy Co$_2$MnSb with direct band gap semiconductor HfIrSb is feasible with both Mn-Sb and Mn-Mn surface terminations. Further, the COHP bonding analysis at the interface suggesting the Mn-Ir bonding (ICOHP:-1.15, -1.46 eV ) at the Mn-Mn/Ir interface is significantly weaker than Sb-Ir bonding (ICOHP:-2.35 eV) at the Mn-Sb/Ir interface. 

The results of tunnel magnetoresistance ratios of the Mn-Mn/Ir and Mn-Sb/Ir interfaces indicate a higher ratio for the latter compared to the earlier. This is due to the significantly less contribution from the minority states in the case of Mn-Sb/Ir interface.  However, experimentally achieved TMR ratio achieved in a heterojunction may be affected by several factors and is limited by the interface quality which is governed by the conditions of growth.

To study the magnetoelectric property and to understand the effect of spin injection bias on the TMR ratio, a transverse electric field in the range of 0.01 to 0.5 V/\AA, in the direction perpendicular to the interfaces, has been included in our calculations. Utilizing the standard two current model given by Julliere, we have calculated the TMR ratio of these heterojunctions under the external electric field. Significantly high TMR ratios have been obtained for these junctions, which is found to remain unaffected by electric field of magnitude up to 0.5 V/\AA. 

Furthermore, we demonstrate that the Co$_2$MnSb/HfIrSb junction displays remarkable strain-sensitive transmission, with a 3-fold increase in majority spin transmission and supression of minority spin-transmission under a bi-axial tensile strain at the Mn-Mn/Ir interface. 

Based on our findings, we predict that a carefully engineered Co$_2$MnSb/HfIrSb junction may have enormous potential for a range of spintronic applications, including magnetic sensors, non-volatile memories, and logic circuits - which awaits experimental validation. 
\section{Acknowledgements}
Authors thank the Director, RRCAT for facilities and encouragement. Authors thank Haiying He for scientific discussions. The computer division of RRCAT, Indore and MTU, USA is thanked for the help in installing and support in smooth running of the codes. JB thanks D. Pandey, R. Dutt, L. Eggart for useful discussions during the work. JB  thanks RRCAT, HBNI and MTU for financial support.
\newpage
\bibliographystyle{apsrev}
\bibliography{Reference}

\end{document}